\documentclass[conference,a4paper]{APSIPA2021}
\usepackage{graphicx}
\usepackage{multirow}
\usepackage{amsmath}
\usepackage[psamsfonts]{amssymb}
\usepackage{amsxtra}
\usepackage{threeparttable}
\usepackage{subfigure}
\usepackage[T1]{fontenc}

\usepackage{fancyhdr}

\fancypagestyle{firststyle} {
    \fancyhf{}
    \fancyhead[L]{Proceedings of 2022 APSIPA Annual Summit and Conference}
    \fancyhead[R]{7-10 November 2022, Chiang Mai, Thailand}

    \fancyfoot[L]{978-616-590-477-3 ©2022 APSIPA}

    \fancyfoot[R]{APSIPA ASC 2022}
}
\fancypagestyle{fa} {
    \fancyhf{}
    \fancyhead[L]{Proceedings of 2022 APSIPA Annual Summit and Conference}
    \fancyhead[R]{7-10 November 2022, Chiang Mai, Thailand}

}

\begin{document}
	
	\title{End-to-end Two-dimensional Sound Source Localization With Ad-hoc Microphone Arrays}
	
	\author{%
		\authorblockN{%
			Yijun Gong, Shupei Liu and
			Xiao-Lei Zhang
		}
		\authorblockA{%
			School of Marine Science and Technology, Northwestern Polytechnical University, China \\
			E-mail: \{gongyj, shupei.liu\}@mail.nwpu.edu.cn, xiaolei.zhang@nwpu.edu.cn  }
	}
	
	\maketitle
	\thispagestyle{firststyle}
	\pagestyle{fa}
	
	\begin{abstract}
		Conventional sound source localization methods are mostly based on a single microphone array that consists of multiple microphones. They are usually formulated as the estimation of the direction of arrival problem. In this paper, we propose a deep-learning-based end-to-end sound source localization method with ad-hoc microphone arrays, where an ad-hoc microphone array is a set of randomly distributed microphone arrays that collaborate with each other. It can produce two-dimensional locations of speakers with only a single microphone per node. Specifically, we divide a targeted indoor space into multiple local areas. We encode each local area by a one-hot code, therefore, the node and speaker locations can be represented by the one-hot codes. Accordingly, the sound source localization problem is formulated as such a classification task of recognizing the one-hot code of the speaker given the one-hot codes of the microphone nodes and their speech recordings. An end-to-end spatial-temporal deep model is designed for the classification problem. It utilizes a spatial-temporal attention architecture with a fusion layer inserted in the middle of the architecture, which is able to handle arbitrarily different numbers of microphone nodes during the model training and test. Experimental results show that the proposed method yields good performance in highly reverberant and noisy environments.
	\end{abstract}
	
	\section{Introduction}
	Sound source localization aims to find the relative position of a sound source to a microphone array by using the multi-channel acoustic signals collected by the microphone array \cite{grumiaux2022survey}. In most cases, sound source localization with a single microphone array could be described as the estimation of the direction of arrival (DOA) of sound sources. The location information is crucial in many applications such as teleconferencing \cite{wang1997voice,hennecke2011towards}, robot audition \cite{valin2004localization,li2016reverberant}, speech separation \cite{chazan2019multi,chen2022end}, speech enhancement \cite{xenaki2018sound} and automatic speech recognition \cite{lee2016dnn}.
	Conventional sound source localization methods mainly include time difference of arrival (TDOA) \cite{knapp1976tdoa}, steered response power with phase transform (SRP-PHAT) \cite{dibiase2000srp}, subspace based multiple signal classification (MUSIC) \cite{schmidt1986music}, etc. Nevertheless, the performance of these conventional methods drops significantly in reverberant and noisy environments. Nowadays, with the development of neural networks, deep-learning-based sound source localization methods e.g. \cite{chen2022end,2019doa,nguyen2020doa}, which show promising performance in the adverse environments, become popular.
	
	However, the aforementioned sound source localization methods are all based on a single microphone array, which mainly obtain merely the directions of sound sources. Moreover, their performance may be unsatisfactory when the sound sources are far away from the microphone array. Ad-hoc microphone arrays may alleviate this problem. An ad-hoc microphone array is a set of randomly distributed microphone nodes with each node containing a single microphone or a microphone array. It reduces the probability of far-field speech processing by grouping the microphone nodes that are close to the sound sources into local microphone arrays. Particularly, deep-learning-based ad-hoc array speech processing shows promising performance in plenty of tasks like speech enhancement \cite{zhang2021deep}, speech separation \cite{yang2022deep}, speech recognition \cite{chen21b_interspeech}, and speaker verification \cite{LiangCGZ21}.
	
	As for the sound source localization problem, an ad-hoc microphone array may be able to estimate the two- or three-dimensional coordinates of sound sources beyond the DOA estimation. Early studies of this direction apply deep learning to distributed microphone arrays \cite{VESPERINI201883}-\cite{kindt20212d}. Specifically, by fixing a given number of microphone nodes on walls, \cite{VESPERINI201883} proposed to predict the 2D coordinates of speaker positions directly. \cite{le2019learning} partitions a room space into grids, and formulated 2D speaker localization as a problem of identifying which spatial grids the speakers belong to. \cite{kindt20212d} applies deep learning based DOA to two distributed microphone nodes respectively, and utilizes triangulation to get the 2D coordinate of a speaker. Although the above methods could cope with the two-dimensional localization task, the methods are limited to small number of microphone nodes such as two nodes, and require matched positions of the microphone nodes in both training and test. These limitations make the methods hard to generalize to mismatched scenarios, and therefore cannot reflect the merits of ad-hoc microphone arrays fully.
	
	In order to promote the advantage of ad-hoc microphone arrays with strong generalization performance, in \cite{Liu2022deep}, a deep-learning-based stage-wise sound source localization with ad-hoc microphone arrays which could contain unfixed number of nodes has been proposed. It first exploits deep neural networks to get the DOA information for each ad-hoc node. Then, it uses a triangulation method to estimate a candidate sound source location from every two nodes. Finally, it conducts clustering on all of the candidate positions for the final accurate location estimation.
	Although the stage-wise sound source localization method has demonstrated its effectiveness in many challenging environments, it consists of many independent modules, which makes it inconvenient to apply in practice. Moreover, some modules are even in unsupervised manner, which do not optimize the localization performance directly. Besides, each microphone node of the ad-hoc array needs to be a conventional microphone array of multiple microphones. The directions and accurate positions of the ad-hoc nodes need to be manually labeled precisely, which is complex and time-consuming.

	\begin{figure}[t]
		\begin{center}
			\begin{minipage}[t]{1\linewidth}
				\includegraphics[width=\textwidth]{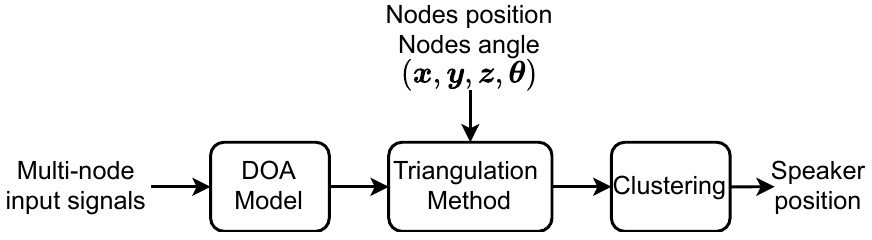}
				\centerline{(a) Stage-wise method \cite{Liu2022deep}}\medskip
				\vspace{0.6cm}
			\end{minipage}
			\centering
			\begin{minipage}[t]{1\linewidth}
				\centering
				\includegraphics[width=68mm,height=2cm]{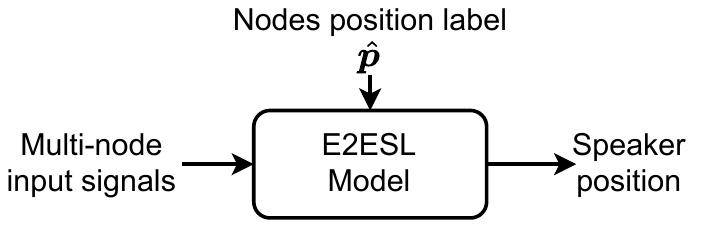}
				\centerline{(b) Proposed end-to-end method}\medskip
			\end{minipage}
		\end{center}
		\caption{Comparison of the architectures of the sound source localization methods with ad-hoc microphone arays.}
		\label{fig:res}
	\end{figure}
	
	To deal with the aforementioned problems, in this paper, we propose an end-to-end sound source localization method (E2ESL) based on ad-hoc microphone arrays to produce the two-dimensional locations of sound sources directly. Similar to \cite{kindt20212d} and \cite{le2019learning}, we also treat the sound source localization problem as a classification problem, which is suitable to the end-to-end formulation of sound source localization. To relax the constraint of the previous studies \cite{kindt20212d,le2019learning} that the node positions in the training and test scenarios have to be the same, we use one-hot codes to encode spatial grids, and use the one-hot codes of the spatial grids where the ad-hoc microphone nodes locate as an input feature of a deep model. The deep model adopts a spatial-temporal attention architecture. It predicts the one-hot codes of the positions of the sound sources. Fig. \ref{fig:res} summarizes the difference between the stage-wise method in \cite{Liu2022deep} and the proposed end-to-end method. We have proved the effectiveness of our method on simulated data with noise and reverberation.
	Our main contributions are as follows:
	\begin{itemize}
		\item  We propose a novel sound source localization method, named E2ESL, which encodes the positions of the ad-hoc microphone nodes as an input feature. This novelty makes E2ESL able to deal with the generalization problem that the locations of the ad-hoc nodes at the training and test scenarios are different.
		\item E2ESL is able to work fine even if each node contains only a single microphone, which is quite different from conventional sound source localization methods where each node has to contain a conventional microphone array, e.g. a linear array. This property may simplify the deployment of E2ESL, since that it does not have to know the direction of the array at each node.
		\item  The deep model of E2ESL applies the attention mechanism to fuse the features produced from the ad-hoc nodes, which is able to handle the case that the numbers of the ad-hoc nodes in training and test are different.
	\end{itemize}
	
	The rest of this paper is organized as follows. Section \ref{sec:PM} introduces the proposed framework and the architecture of the end-to-end deep model. Experimental results are shown in Section \ref{sec:exp}. Finally, Section \ref{sec:con} draws a conclusion.

	\section{Proposed method}
	\label{sec:PM}
	
	\subsection{Framework}
	\label{ssec:sub2-1}
	We formulate the E2ESL method with ad-hoc microphone arrays as a classification problem. Given a room of size $\{X,Y,Z\}$. Suppose there are $N$ randomly distributed nodes in the room, where each node contains a single microphone. Suppose the cartesian coordinate of the $n$th node in the room is $\mathbf{p}_n=\{x_n,y_n,z_n\}$, and $\mathbf{P} =\{\mathbf{p}_n | \forall n = 1,\ldots,N\}$ represents the set of the coordinates of the nodes. Suppose the coordinate of a speaker is $\mathbf{p}_{\mathrm{spkr}}=\{x_{\mathrm{spkr}},y_{\mathrm{spkr}},z_{\mathrm{spkr}}\}$. Here we only consider the 2-dimensional (2D) coordinate estimation of the speaker, though the proposed framework is not limited to the 2D coordinate estimation. In the following, the height coordinates $Z$, $z_n$, and $z_{\mathrm{spkr}}$ are omitted. Suppose each node contains a single microphone, though the proposed method is not limited to this assumption. Suppose the speech signal recordings from the $N$ nodes are $\mathbf{S}= \{\mathbf{s}_n(t)| \forall t = 1,\ldots, T,\ \forall n = 1,\dots,N\}$, where $\mathbf{s}_n(t)$ denotes the speech signal recorded by the $n$th node at the $t$th frame.
	
	\begin{figure}[t]
		\begin{center}
			\includegraphics[width=0.85\linewidth]{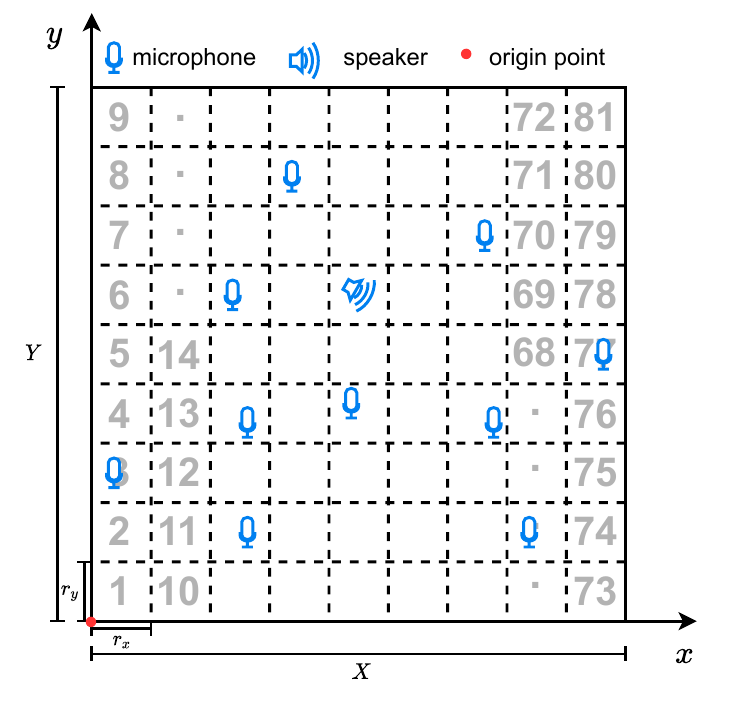}
		\end{center}
		\caption{Formulation of the E2ESL method with ad-hoc microphone arrays as a classification problem. The symbols $r_x$ and $r_y$ in the figure are the resolutions along the $x$ and $y$ axes which determine the size of each local area.}
		\label{fig:local area}
	\end{figure}
	
	We establish a cartesian coordinate system for the room, where we select a corner of the room as the origin. We first split the room uniformly into $M$ local areas, as shown in Fig. \ref{fig:local area}. Then, we assign each local area an index. Given the size of the local area denoted as $\{r_x, r_y\}$, $M$ could be obtained by:
	\begin{align}
	M = M_x  M_y = \frac{X}{r_x} \frac{Y}{r_y} .
	\label{equ:blocknum}
	\end{align}
	The coordinates that the $m$th rectangle local area covers are in the region surrounded by the four corners of the rectangle:
	\begin{align}
	&b^{mx}_{start} = ((m\!-\!1)//M_y)\!\times\!r_x \\
	&b^{mx}_{end} = (((m\!-\!1)//M_y)\!+\!1)\!\times\!r_x \\
	&b^{my}_{start} = ((m\!-\!1)\%M_y)\!\times\!r_y \\
	&b^{my}_{end} = (((m\!-\!1)\%M_y)\!+\!1)\!\times\!r_y
	\end{align}
	where the symbols ``$//$'' and  ``$\%$'' are the floor division and modulo operators respectively. Given the above formulation, the position of an ad-hoc node $\mathbf{p}_n$ is simplified to the index of the local area where the node is located, which is denoted as $\hat{p}_n \in \{1,...,M\}$, so as to the sound sources.
	
	E2ESL takes speech recordings of the ad-hoc array $S$ and the indexes of the positions of the ad-hoc nodes $\hat{P} = \{\hat{p}_1,\ldots, \hat{p}_N\}$ as the input to estimate the index of the local area where the speaker locates $\hat{p}_{\mathrm{spk}} \in \{1,...,M\}$:
	\begin{equation}
	\hat{p}_{\mathrm{spk}} = f(\hat{P}, S)
	\end{equation}
	where $f(\cdot)$ is the deep model described in the next subsections. Finally, the center of the predicted local area is regarded as the coordinates of the estimated speaker position.
	
	\subsection{Implementation}

	The general pipeline of the proposed E2ESL is given in Fig. \ref{fig:structure}. It first takes the one-hot code and short-term Fourier transform (STFT) acoustic feature of the ad-hoc node as the inputs of two encoders respectively. Then, it concatenates the outputs of the two encoders, and takes the concatenated feature as the input of a spatial-temporal attention network. The attention network conducts cross-channel attention, channel fusion, and cross-frame attention successively. Finally, it averages the output of the spatial-temporal attention network along the time axis for the estimation of the one-hot code of the local area of the speaker. The details of the proposed E2ESL are described as follow:

	\begin{figure}[t]
		\begin{center}
			\hspace{-4.1mm}
			\includegraphics[width=0.98\linewidth]{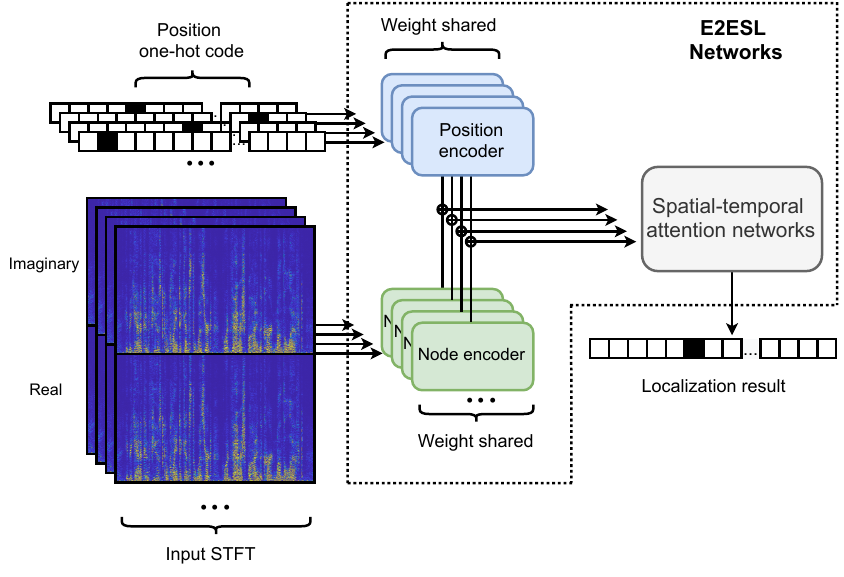}
		\end{center}
		\caption{Diagram of the end-to-end sound source localization algorithm.}
		\label{fig:structure}
	\end{figure}
	
	\subsubsection{One-hot encoding for the local areas}
	\label{ssec:sub2-2}
	Because the position indexes $\hat{P}$ of the ad-hoc nodes are not suitable to be a feature, here we encode the indexes by the one-hot codes, though other coding methods can be applied as well. For a local area of index $m$, the one-hot code is an $M$-dimensional sparse vector with the $m$th element set to 1 and the other elements set to 0.
	By doing this, we can represent the positions of the ad-hoc nodes and speakers as the one-hot vectors. We denote the one-hot codes of the ad-hoc nodes as $\mathbf{U} = \{\mathbf{u}_n| \forall n = 1,\ldots,N \}$, and the one-hot code of the speaker as $\mathbf{l}$.

	\subsubsection{Acoustic feature extraction}
	We extract time-frequency spectrograms from the multi-channel signals of the ad-hoc microphone array by STFT, and then concatenate the real and imaginary parts as the acoustic feature:
	\begin{align}
	\mathbf{\hat{S}} = &{\rm Concat}\left({\rm Imag}({\rm STFT}(\mathbf{\mathbf{S}})),  {\rm Real}({\rm STFT}(\mathbf{S}))\right)  \ \nonumber\\
	=& \{\mathbf{\hat{s}}_n(f,t)  |  \forall f=1,\ldots,2F, \forall  n=1,\ldots,N, \forall t=1,\ldots,T\}\label{equ:stft}
	\end{align}
	where $\mathbf{{s}}_n(f,t)$ denotes the concatenated feature of the $n$th node at time $t$ and frequency $f$.
	
	\subsubsection{Encoders}
	We feed the acoustic feature $\mathbf{\hat{S}}$ and the one-hot vectors $\mathbf{U}$ of the ad-hoc nodes to a \textit{node encoder} and a \textit{position encoder} respectively:
	\begin{align}
	\mathbf{\tilde{s}}_n = &{\rm NodeEncoder}(\mathbf{\hat{s}}_n),\ \forall n=1,\ldots,N \label{equ:NE}\\
	\mathbf{\tilde{u}}_n = &{\rm PositionEncoder}(\mathbf{u}_n),\ \forall n=1,\ldots,N \label{equ:PE}
	\end{align}
	where $\mathbf{\tilde{s}}_n$ and $\mathbf{\tilde{u}}_n$ are the \textit{acoustic embedding} and \textit{position embedding} of the $n$th ad-hoc node. The two encoders, which consist of multiple linear layers, are shared by all ad-hoc nodes.
	
	Then, for each ad-hoc node, the position embedding $\mathbf{\tilde{u}}_n$ is first duplicated by $T$ times, and then concatenated with  $\mathbf{\tilde{s}}_n$:
	\begin{align}
	\mathbf{e}_n =& {\rm Concat}({\rm Duplicate}(\mathbf{\tilde{u}}_n),  \mathbf{\tilde{s}}_n),\ \forall n=1,\ldots,N.
	\end{align}
	We denote $\mathbf{E}=\{\mathbf{e}_n \in \mathbb{R}^{ D\times T}| \forall n = 1,\ldots, N\}$ where $D$ is the dimension of $\mathbf{e}_n$.

	\subsubsection{Spatial-temporal attention network}

\begin{figure}[htb]
		\begin{center}
			\subfigure[\label{attention}]{
				\includegraphics[width=0.98\linewidth]{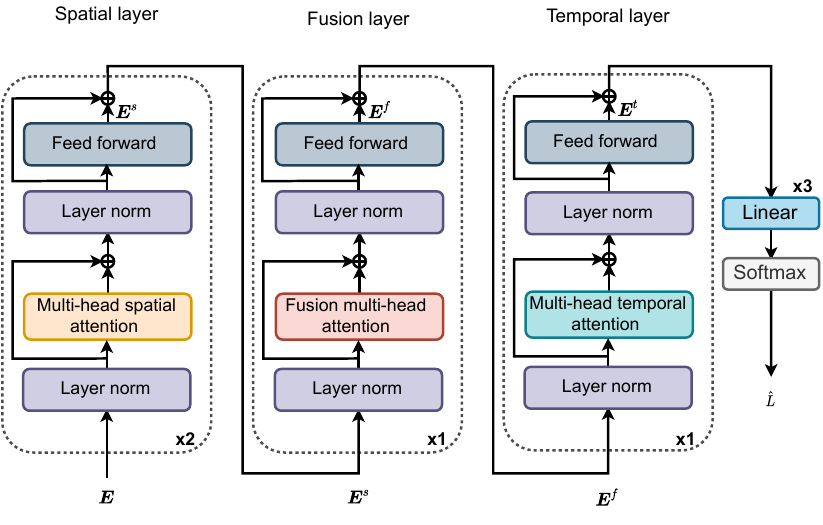}
				
			}
			\\
			
			\vspace{-1mm}
			\subfigure[\label{sa}]{
				\hspace{-6mm}	
				\includegraphics[width=0.9in]{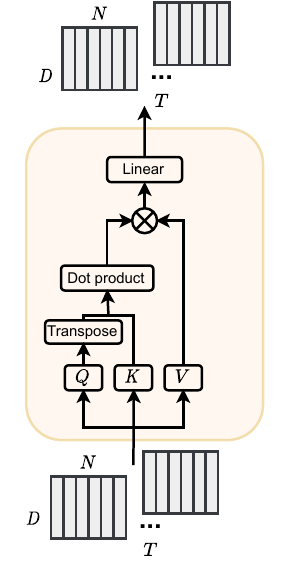}	
			}
			\hspace{0.5mm}	
			\vspace{2mm}
			\subfigure[\label{fa}]{
				\includegraphics[width=0.9in]{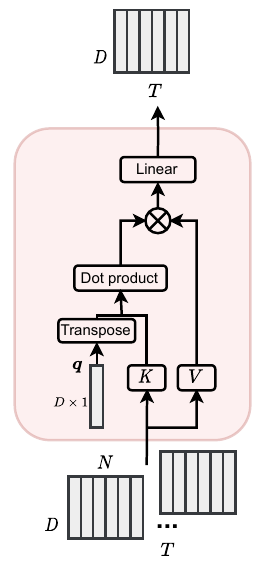}	
			}
			\hspace{-1mm}	
			\vspace{-1mm}
			\subfigure[\label{ta}]{	
				\includegraphics[width=0.9in]{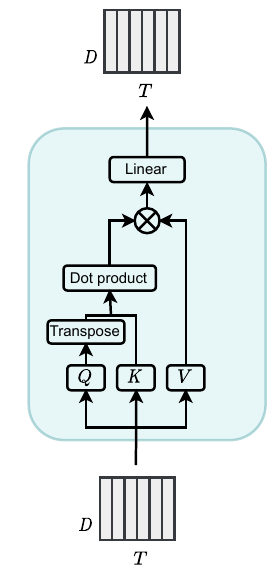}	
			}
			\hspace{-1mm}	
			\vspace{-2mm}
		\end{center}
		\caption{Spatial-temporal attention network.
			(a) General architecture. (b) Spatial layer. (c) Fusion layer. (d) Temporal layer.}
		\vspace{-2mm}
	\end{figure}

	As shown in Fig. \ref{attention}, the spatial-temporal attention network consists of three main parts: two stacked spatial layers, a fusion layer and a temporal layer. Each layer consists a multi-head attention sub-layer and a feed forward sub-layer. Besides, the layer normalization and residual connection are also applied.
	
	\textbf{Spatial layer:}
	We first feed $\mathbf{E}$ into the spatial layer frame by frame for the cross-channel attention, which learns a new multi-channel feature that contains the global spatial connection between the ad-hoc nodes, denoted as $\mathbf{E}^s \in \mathbb{R}^{ T\times D \times N} $:
	\begin{align}
	\mathbf{E}^s = &{\rm FFN}(\mathbf{E}+{\rm MHA}(\mathbf{E}) ) \nonumber \\
	= &{\rm Linear}({\rm Linear}(\mathbf{E}+{\rm MHA}(\mathbf{E}))) \label{eq:ffn}
	\end{align}
	where ${\rm FFN}$ and ${\rm MHA}$ denote the feed-forward module and multi-head attention networks. FFN consists of two linear layers as shown in (\ref{eq:ffn}), and ${\rm MHA}$ is defined as:
	\begin{equation}
	{\rm MHA}(\mathbf{E})= {\rm Concat}(\mathbf{O}^s_1,\ldots,\mathbf{O}^s_h,\ldots, \mathbf{O}^s_H)\mathbf{W}^o \label{eq:mha2}
	\end{equation}
	where $H$ denotes the number of heads, $\mathbf{W}^o \in \mathbb{R}^{D \times D}$ is the output projection matrix, and $\mathbf{O}^s_h$ with $h=1,...H$ is the output of each attention head:
	
	\begin{equation}
	\mathbf{O}^s_h = {\rm Softmax} \left(\frac{(\mathbf{W}^q_h\mathbf{E})^T(\mathbf{W}^k_h\mathbf{E})}{\sqrt{D/H}}\right)(\mathbf{W}^v_h\mathbf{E}),\ \forall h=1,...H\label{eq:attention}
	\end{equation}
	where $\mathbf{W}^q_h \in \mathbb{R}^{ D/H \times N}$, $\mathbf{W}^k_h\in \mathbb{R}^{ D/H \times N}$, and $\mathbf{W}^v_h\in \mathbb{R}^{ D/H \times N}$ are learnable projection matrices.

	\textbf{Fusion layer:} After conducting the multi-head attention across channels by the spatial layer, the multi-channel features produced from the spatial layer are fused into a single-channel feature through the fusion layer. As shown in Fig. \ref{fa}, the attention module in the fusion layer is the same as the standard multi-head attention in the spatial layer except that $\mathbf{W}^q_h\mathbf{E^s}$ is replaced by a learnable vector $\mathbf{q} \in \mathbb{R}^{D \times 1}$. The output of the fusion layer $\mathbf{E}^f$
	\begin{align}
	\mathbf{E}^f = &{\rm FFN}(\mathbf{E}^s+{\rm FMHA}(\mathbf{E}^s) )
	\end{align}
	where ${\rm FMHA}$ is the fusion multi-head attention:
	\begin{align}
	{\rm FMHA}(\mathbf{E}^s) = &{\rm Concat}(\mathbf{O}^f_1,\ldots,\mathbf{O}^f_h,\ldots, \mathbf{O}^f_H)\mathbf{W}^o
	\end{align}
	with $\mathbf{O}^f_h$ calculated by:
	\begin{align}
	\mathbf{O}^f_h = &{\rm Softmax} \left(\frac{\mathbf{q}^T(\mathbf{W}^k_h\mathbf{E}^s)}{\sqrt{D/H}}\right)\left(\mathbf{W}^v_h\mathbf{E}^s\right),\ \forall h=1,\ldots,H
	\end{align}
	
	\textbf{Temporal layer:}
	We feed $\mathbf{E}^f$ into the temporal layer for the cross-frame attention. The architecture of the multi-head attention module in the temporal layer is the same as that in the spatial layer. The output of the temporal layer $\mathbf{E}^t \in \mathbb{R}^{D \times T}$ is calculated by:
	\begin{align}
	\mathbf{E}^t = &{\rm FFN}(\mathbf{E}^f+{\rm MHA}(\mathbf{E}^f) ).
	\end{align}
	
	Finally, we obtain the classification result $\mathbf{\hat{L}} \in \mathbb{R}^{T\times M}$ through a stack of linear layers and a softmax output layer:
	\begin{equation}
	\hat{\mathbf{L}} = {\rm Softmax}({\rm Linear}(\mathbf{E}^t)).
	\end{equation}
	
	\subsubsection{Postprocessing and coordinate estimation of speakers}
	
	We average $\hat{\mathbf{L}}$ over the time dimension for an estimation of the one-hot code of the speaker $\hat{\mathbf{l}}\in \mathbb{R}^{M}$. We conduct a hard decision to $\hat{\mathbf{l}}\in \mathbb{R}^{M}$, which selects the index of the class with the highest probability as the estimated local area $\hat{l}$. The estimated coordinate of the speaker $\tilde{\mathbf{p}}_{\mathrm{spkr}} = \{\tilde{x}_{\mathrm{spkr}},\tilde{y}_{\mathrm{spkr}}\}$ is obtained as the center of the local area:
	\begin{align}
	\tilde{x}_{\mathrm{spkr}} = &\frac{1}{2}[(((\hat{l}\!-\!1)//M_y)\!+\!1)\!\times\!r_x+((\hat{l}\!-\!1)//M_y)\!\times\!r_x] \\
	\tilde{y}_{\mathrm{spkr}} = &\frac{1}{2}[(((\hat{l}\!-\!1)\%M_y)\!+\!1)\!\times\!r_y+((\hat{l}\!-\!1)\%M_y)\!\times\!r_y].
	\end{align}

	\section{Experiments}
	\label{sec:exp}
	\subsection{Experimental setup}
	\label{ssec:expsetup}
	We conducted an experiment on a simulated dataset. The source speech is from the TIMIT corpus \cite{1993timit}, which consists of 462 speakers in the training set and 168 speakers in the test set. We first removed silent segments of the speech recordings by voice activity detection. Then, we cut each recording into segments of two-second long, and discard the segments that are shorter than two seconds. For each simulated speech recording for training, the source speech is a randomly chosen segment from a speaker in the training set of TIMIT, so as to the simulated test data.
	The image source model \cite{1979image}\footnote{https://github.com/ehabets/RIR-Generator} is exploited to generate reverberant environments. Diffused white noise is added by the noise generators.\footnote{https://www.audiolabs-erlangen.de/fau/professor/habets/software/noise-generators} The reverberation time T$_{60}$ for each sample is chosen randomly from a range of $[0.3, 1.0]$ second. The SNR at the position of the speaker is randomly selected from a range of  $[20,50]$ dB.

	For the simulated training data, we generated 10 rooms with different sizes. Each room was divided into 225 local areas of equal size. We randomly selected 36 positions in each local area as the speaker position. For each speaker position, we randomly picked a segment from the training set of TIMIT. Eventually, we generated 81,000 source samples for training with each sample being 2 seconds long. The duration of the training set is 45 hours long. For each sample, we generated an ad-hoc microphone array of 30 microphone nodes, where the 30 nodes were randomly put into 30 local areas. The microphone position was randomly chosen inside the local area.

	The TIMIT test set was partitioned into two parts, which were used to construct the simulated validation data and test data respectively. For the simulated test data, we simulated 5 rooms, each of which also contains 225 local areas. We randomly selected 5 speaker positions in each local area, each of which corresponds to a randomly selected test sample. The other settings were the same as the training set. The room sizes of the simulated training and test data are listed in Table \ref{table:room}, which as we can see are different.
	
	\begin{table}[t]
		\begin{center}
			\begin{threeparttable}
				\caption{Room size in training and test set}\label{table:room}
				
				\begin{tabular}{|c|c|c|}
					\hline
					\textbf{Dataset} &  \textbf{Room} & \textbf{Room size} \\
					\hline
					\multirow{10}{*}{\centering Training set}
					&Room1&$5\times 7\times 3$ m\\
					\cline{2-3}
					&Room2&$10\times 10\times 3$ m\\
					\cline{2-3}
					&Room3&$7\times 9\times 3$ m\\
					\cline{2-3}
					&Room4&$8\times 8\times 3$ m\\
					\cline{2-3}
					&Room5&$8\times 6\times 3$ m\\
					\cline{2-3}
					&Room6&$5\times 5\times 3$ m\\
					\cline{2-3}
					&Room7&$7\times 4\times 3$ m\\
					\cline{2-3}
					&Room8&$6\times 7\times 3$ m\\
					\cline{2-3}
					&Room9&$5\times 4\times 3$ m\\
					\cline{2-3}
					&Room10&$9\times 8\times 3$ m\\

					\hline
					\multirow{5}{*}{\centering Test set}
					&Room1&$5\times 6\times 3$ m\\
					\cline{2-3}
					&Room2&$7\times 8\times 3$ m\\
					\cline{2-3}
					&Room3&$5\times 8\times 3$ m\\
					\cline{2-3}
					&Room4&$6\times 6\times 3$ m\\
					\cline{2-3}
					&Room5&$6\times 9\times 3$ m\\
					
					\hline
				\end{tabular}
				
			\end{threeparttable}
		\end{center}
	\end{table}
	
	\begin{table}[t]
		\begin{center}
			\begin{threeparttable}
				\caption{Model structure of the proposed E2ESL.}\label{table:network}
				
				\begin{tabular}{|c|c|c|c|}
					\hline
					\textbf{Module} &  \textbf{Network} & \textbf{Input size} & \textbf{Output size}\\
					
					\hline
					\multirow{2}{*}{\centering Position  encoder}
					&Linear&$M$&$512$\\
					\cline{2-4}
					&Linear&$512$&$256$\\
					
					\hline
					\multirow{3}{*}{\centering Node  encoder}
					&Linear&$512$&$1024$\\
					\cline{2-4}
					&Linear&$1024$&$512$\\
					\cline{2-4}
					&Linear&$512$&$256$\\
					
					\hline
					\multirow{2}{*}{\centering Spatial layer}
					&Self-attention\tnote{*}&$512$&$512$\\
					\cline{2-4}
					&Feed forward&$512$&$512$\\
					\cline{2-4}

					\hline
					\multirow{2}{*}{\centering Fusion layer}
					&Self-attention\tnote{*}&$512$&$512$\\
					\cline{2-4}
					&Feed forward&$512$&$512$\\
					\cline{2-4}
					
					\hline
					\multirow{2}{*}{\centering Temporal layer}
					&Self-attention\tnote{*}&$512$&$512$\\
					\cline{2-4}
					&Feed forward&$512$&$512$\\
					\cline{2-4}
					
					\hline
					\multirow{3}{*}{\centering Output layer}
					&Linear&$512$&$1024$\\
					\cline{2-4}
					&Linear&$1024$&$1024$\\
					\cline{2-4}
					&Linear&$1024$&$M$\\
					\cline{1-4}
				\end{tabular}
				\begin{tablenotes}
					\item[*] We used 4 heads attention in our experiments.
				\end{tablenotes}
			\end{threeparttable}
		\end{center}
	\end{table}
	
	The sampling rate of the simulated data is 16KHz. We conducted 512 points STFT for each frame with a frame overlap of $50\%$. The details of the network structure are summarized in Table \ref{table:network}. For each layer, ReLU was applied as the activation function. In the training process, we used minimum cross entropy as the loss function, and used Adam as the optimizer.
	
	We used mean absolute error (MAE) to evaluate the performance of the proposed algorithm:
	\begin{equation}
	MAE = \frac{1}{I}\sum_{i=1}^{I}\sqrt{(y_{\mathrm{spkr},i}-\tilde{y}_{\mathrm{spkr},i})^2+(x_{\mathrm{spkr},i}-\tilde{x}_{\mathrm{spkr},i})^2}
	\end{equation}
	where $I$ is the number of the test samples. The lower the MAE is, the better the performance is.

	\subsection{Main results}
	
	We trained E2ESL by the 45 hours training data with 30 nodes, and tested the model on the 5 test rooms with the number of the ad-hoc nodes varying from 20 to 30. When the number of the ad-hoc nodes in a test scenario is less than 30, the nodes are randomly selected from the 30 nodes.
	
	The results are shown in Table \ref{table:nodes_exp}. From the table, we see that, with the increasing of the number of nodes, MAE decreases in all 5 test rooms. For example, the average MAE in the scenario with 30 nodes is 6 cm lower than that with 20 nodes.
	
	Note that the reason why we do not compare with other sound source location methods, e.g. \cite{Liu2022deep}, is that our test scenario where each ad-hoc node contains only a single microphone makes the methods unapplicable.
	\begin{table}[t]
		\begin{center}
			\begin{threeparttable}
				\caption{MAE (in meters) of the proposed method on the simulated test data with different numbers of ad-hoc nodes. }\label{table:nodes_exp}
				\tabcolsep 5pt
				\begin{tabular}{|c|c|c|c|c|c|c|}
					\hline
					\textbf{Nodes}&\textbf{Room1}&\textbf{Room2}&\textbf{Room3}&\textbf{Room4}&\textbf{Room5}&\textbf{Average} \\
					\hline
					30 &0.2271&\textbf{0.2838}&\textbf{0.2785}&0.2292&\textbf{0.3127}&\textbf{0.2662}\\
					\hline
					29 &\textbf{0.2208}&0.2873&0.2804&\textbf{0.2282}&0.3168&0.2667\\
					\hline
					28 &0.2221&0.2857&0.2831&0.2350&0.3227&0.2697\\
					\hline
					27 &0.2221&0.2917&0.2907&0.2360&0.3278&0.2736\\
					\hline
					26 &0.2336&0.3012&0.2986&0.2413&0.3357&0.2820\\
					\hline
					25 &0.2454&0.3073&0.3024&0.2467&0.3391&0.2881\\
					\hline
					24 &0.2487&0.3120&0.3167&0.2519&0.3432&0.2945\\
					\hline
					23 &0.2567&0.3187&0.3174&0.2572&0.3577&0.3015\\
					\hline
					22 &0.2575&0.3123&0.3258&0.2685&0.3738&0.3075\\
					\hline
					21 &0.2703&0.3337&0.3320&0.2736&0.3837&0.3186\\
					\hline
					20 &0.2735&0.3327&0.3437&0.2887&0.4075&0.3292\\
					\hline
					
				\end{tabular}
			\end{threeparttable}
		\end{center}
	\end{table}
	\subsection{Effect of the size of the local area}
	It is obvious that the size of the local area affect the performance of E2ESL. To exam its effect, we split a room
	into 100 or 64 local areas with an equal size respectively. We compared the E2ESL models trained with 225, 100 and 64 local areas in room1 of the simulated test data. Table \ref{table:blocks_exp} shows the experimental result where ACC denotes the classification accuracy. From the table, we find interestingly that, when the resolution of the room is improved from 64 to 225 local areas, although the classification accuracy decreases from 70.01\% to 52.44\%, the localization performance by MAE is improved from 0.3285 meters to 0.2271 meters.

	\begin{table}[t]
		\begin{center}
			\begin{threeparttable}
				\caption{Results on test room1 with different local area numbers}\label{table:blocks_exp}
				\tabcolsep 4.5pt
				\begin{tabular}{|c|c|c|c|c|c|}
					\hline
					\multicolumn{2}{|c|}{\textbf{225 local areas}}&\multicolumn{2}{|c|}{\textbf{100 local areas}}&\multicolumn{2}{|c|}{\textbf{64 local areas}}\\
					\hline
					MAE (m)&ACC (\%)&	MAE (m)&ACC (\%)&	MAE (m)&ACC (\%)\\
					\hline
					0.2271&52.44&0.2828&62.25&0.3285&70.01\\
					\hline
				\end{tabular}
			\end{threeparttable}
		\end{center}
	\end{table}
	
	\section{Conclusion}
	\label{sec:con}
	In this paper, we have proposed an end-to-end sound source localization method with ad-hoc microphone arrays where each ad-hoc node contains a single microphone. With a room splitting strategy and one-hot encoding, we leverage location information into the classification based end-to-end sound source localization system, which only needs rough indexes of the local areas as the position features of the ad-hoc nodes. A spatial-temporal attention network is utilized to capture the global information of speakers and acoustic environments in the dimensions of both space and time. We demonstrated the effectiveness of the proposed algorithm in the simulated data with additive noise and strong reverberation, given the situation where each ad-hoc node contains only a single microphone.
	
	\section{Acknowledgement}

This work was supported in part by National Science Foundation of China under Grant No. 62176211, in part by Project of the Science, Technology, and Innovation Commission of Shenzhen Municipality under grant No. JSGG20210802152546026 and JCYJ20210324143006016.
	
	\bibliographystyle{IEEEbib}
	\bibliography{apsipa}

\end{document}